\documentclass[sn-mathphys]{sn-jnl}

\jyear{2021}%

\theoremstyle{thmstyleone}%
\theoremstyle{thmstyletwo}%

\theoremstyle{thmstylethree}%

\begin{document}

\title{Aggregate Effects of Proliferating LEO Objects and Implications for Astronomical Data Lost in the Noise}

\author*[1,2]{\fnm{John~C.} \sur{Barentine}}\email{john@darkskyconsulting.com}
\author[3]{\fnm{Aparna} \sur{Venkatesan}}\email{avenkatesan@usfca.edu}
\author[4]{\fnm{Jessica} \sur{Heim}}\email{Jessica.Heim@usq.edu.au}
\author[5]{\fnm{James} \sur{Lowenthal}}\email{jlowenth@smith.edu}
\author[6,7]{\fnm{Miroslav} \sur{Kocifaj}}\email{miroslav.kocifaj@savba.sk}
\author[8]{\fnm{Salvador} \sur{Bará}}\email{salva.bara@usc.gal}

\affil*[1]{\orgname{Dark Sky Consulting, LLC}, \orgaddress{\street{9420 E Golf Links Rd Ste 108 PMB 237}, \city{Tucson}, \state{AZ}, \postcode{85730-1317}~\country{USA}}}

\affil[2]{\orgdiv{Consortium for Dark Sky Studies}, \orgname{University of Utah}, \orgaddress{\street{375 S 1530 E, RM 235 ARCH}, \city{Salt Lake City}, \state{UT} \postcode{84112-0730}~\country{USA}}}

\affil[3]{\orgdiv{Department of Physics and Astronomy}, \orgname{University of San Francisco}, \orgaddress{\street{2130 Fulton Street}, \city{San Francisco}, \state{CA}~\postcode{94117} \country{USA}}}

\affil[4]{\orgdiv{Centre for Astrophysics}, \orgname{University of Southern Queensland}, \orgaddress{\street{West Street}, \city{Toowoomba}, \state{Queensland} \postcode{4350} \country{Australia}}}

\affil[5]{\orgdiv{Five College Astronomy Department}, \orgname{Smith College}, \orgaddress{\street{44 College Lane}, \city{Northampton}, \state{MA}~\postcode{01063} \country{USA}}}

\affil[6]{\orgdiv{Institute of Construction and Architecture}, \orgname{Slovak Academy of Sciences}, \orgaddress{\street{Dúbravská cesta 9}, \postcode{845 03} \city{Bratislava}, \country{Slovakia}}}

\affil[7]{\orgdiv{Department of Experimental Physics}, \orgname{Comenius University}, \orgaddress{\street{Mlynská dolina}, \postcode{842 48} \city{Bratislava}, \country{Slovakia}}}

\affil[8]{\orgname{Agrupación Astronómica Coruñesa Ío}, \orgaddress{\street{A Coruña}, \city{Galicia}, \country{Spain}}}


\abstract{The rising population of artificial satellites and associated debris in low-altitude orbits is increasing the overall brightness of the night sky, threatening ground-based astronomy as well as a diversity of stakeholders and ecosystems reliant on dark skies. We present calculations of the potentially large rise in global sky brightness from space objects, including qualitative and quantitative assessments of how professional astronomy may be affected. Debris proliferation is of special concern: since all log-decades in debris size contribute approximately the same amount of night sky radiance, debris-generating events are expected to lead to a rapid rise in night sky brightness along with serious collision risks for satellites from centimetre-sized objects. This will lead to loss of astronomical data and diminish opportunities for ground-based discoveries as faint astrophysical signals become increasingly lost in the noise. Lastly, we discuss the broader consequences of brighter skies for a range of sky constituencies, equity/inclusion and accessibility for Earth- and space-based science, and cultural sky traditions. Space and dark skies represent an intangible heritage that deserves intentional preservation and safeguarding for future generations.}


\maketitle

\section{Space objects in LEO are proliferating rapidly}\label{sec1}

Orbital space near the Earth has been transformed radically since the launch of the first artificial satellite in 1957. The number of functional satellites in low-Earth orbit (LEO) has more than doubled since early 2019 due to the advent of large groups of satellites informally known as ‘megaconstellations’. As LEO becomes an increasingly congested space, the risk of collisions between and among objects increases exponentially and risks an uncontrolled chain reaction of debris-generating events.

In only three years, satellite megaconstellations have become an increasingly serious threat to astronomy. We are witnessing a dramatic, fundamental, and perhaps semi-permanent transformation of the night sky without historical precedent and with limited oversight. The number of satellites planned for launch in the 2020s and beyond is enormous,~\cite{Lawrence2022} driven primarily by private companies motivated by profit. Their potential effects on astronomy are in need of more basic data on the varying brightnesses of satellites in orbit. This is also beginning to be studied through simulations, data processing solutions and calculations of impacts;~\cite{Walker2020a,Walker2020b,Hall2021,WalkerBenvenuti2022} see also the ``Trailblazer'' open data repository.~\cite{Rawls2022} Nevertheless, we fear that faint astrophysical signals will become increasingly lost in the noise due to satellite megaconstellations.

\section{LEO crowding is changing the nature of the space environment}\label{sec2}

Direct illumination by sunlight of functional satellites, failed satellites, leftover launch hardware, and debris fragments (collectively, “space objects''), makes them visible as streaks or trails in astronomical optical and infrared images, which can compromise scientific data.~\cite{Hall2021,McDowell2020,HainautWilliams2020,Ragazzoni2020,Mroz2022,LawlerBoleyRein2021} Myriad smaller objects contribute to elevating the diffuse brightness of the night sky. Kocifaj et al. estimated that, even before the megaconstellation era began, space objects already contributed additional light at the zenith amounting to as much as 10\% of an assumed natural background level.~\cite{Kocifaj2021}

Large objects such as intact satellites make a small but non-negligible contribution to diffuse night sky brightness (NSB).~\cite{Bassa2022} A greater concern is the generation of small debris. As of mid-2022, the number of objects larger than 10 cm in size orbiting the Earth was estimated at around 36,500.~\cite{ESA2022} Below that size scale there is little publicly available information on the population of debris. The number of centimetre-sized objects that could seriously damage satellites in collisions is probably around one million. 

These small objects also disproportionately contribute to rising diffuse NSB. The cumulative numbers $N$($>D$) of space objects with a lower size limit ($D$) ranging from one micron to roughly five metres in diameter adopted by Kocifaj et al.~\cite{Kocifaj2021} — representing the latest and most complete data available to civilian scholars — implies that all log-decades in object size contribute approximately the same amount of night sky radiance. Therefore, a rapid rise in NSB is probable if space debris proliferates significantly. That remains true even if the rate of new launches slows or stops altogether, as the population of objects with smaller sizes will likely increase as a result of newly generated debris from existing satellites. There are no known effective mitigations for the problem of elevated night sky brightness other than drastic reductions in satellite launches or satellite brightness. 

Satellites also pose a threat to astronomy outside the visual and near-infrared bands.~\cite{McDowell2020} Radio astronomy is vulnerable to the direct and indirect emissions of radio energy from satellites and in particular to out-of-band transmissions and sidelobes. Direct illumination of radio telescopes by satellites transmitting to the ground could damage or destroy sensitive radio detectors.~\cite{Walker2020b}

Some scholars identify an environmental continuum between Earth and space that calls for a reconsideration of space ‘sustainability’ and define near-Earth space as part of the human environment.~\cite{Yap2022} A new ‘space environmentalism’ framework has been suggested to manage outer space sustainably and equitably.~\cite{Lawrence2022,Hall2021}

Despite these efforts, many areas of astronomical research will be increasingly affected in a future in which LEO is ever more crowded with satellites. It is unrealistic and economically infeasible to move astronomy exclusively to space, and space telescopes can also be affected by orbital debris and satellite trails in images.~\cite{Kruk2021} Some of the effects on astronomical images can be mitigated with software;~\cite{Tyson2020,Hasan2021} however, this approach is expensive and imperfect. Not all effects, such as detector crosstalk during the passage of very bright objects through the field of view, can be removed. These destructive events risk opportunity loss that may impact scientific productivity of facilities and impede discovery. An overall increase in diffuse NSB requires longer exposure times to reach particular detection thresholds, which in turn increases the likelihood that streaks from resolved objects will affect images, and it increases the cost of data collection as we detail below. Unintended radio interference from satellites can hamper or prevent radio astronomy data collection. 

These outcomes imply a diminution of future opportunities for discovery from both the ground and LEO. We next consider their associated costs. 

\section{The tangible costs of rising global night sky brightness}\label{sec3}

For over a century, new ground-based astronomical observatories have been established in increasingly remote places, in part to evade the influence of terrestrial skyglow. Even when such sites are identified and developed, the threat of skyglow remains ever-present.~\cite{Green2022,Falchi2022} Increased NSB resulting from proliferating space objects is a fundamentally new challenge to astronomy. In choosing new observatory sites, one cannot simply look for more distant locations because increasing global sky brightness will be experienced planet-wide. Here we attempt to quantify what this means for the future of terrestrial and LEO-based astronomy.

\subsection{Loss of data and discovery for professional astronomy}

As NSB rises, the exposure time required to reach any particular signal-to-noise ratio (S/N) rises concomitantly. In cases where exposure times are held fixed, such as in areal sky surveys, a brighter night sky corresponds to a brighter detection limit. As a result, fainter objects will be missed, which will directly diminish the pace and impact of astronomical discovery. It is impossible to put a reliable monetary cost estimate on the loss of opportunity, particularly if we miss rare astrophysical phenomena because satellites interfered with observations. An example with distinct and potentially severe social consequences is the detection of near-Earth objects (NEOs) that represent a high risk of colliding with our planet. For example, the Chelyabinsk bolide – the largest known natural object to have entered Earth's atmosphere since the 1908 Tunguska event – was undiscovered at the time of its entry into the Earth’s atmosphere in 2013 in part due to its position on the sky, near the Sun, in the days and hours before impact.~\cite{Borovicka2013} Hazardous NEOs that sky surveys may fail to detect often first appear in our skies in the twilight hours around sunrise and sunset, times when satellites and space debris are most likely to interfere with observations.~\cite{Walker2020a}

This could have profound consequences for high-profile terrestrial facilities in the coming decades. For example, the Vera Rubin Observatory estimates that if the SpaceX Starlink constellation achieved its full design buildout of 42,000 satellites, as many as 30\% of all Legacy Survey of Space and Time (LSST) images would contain at least one satellite trail.~\cite{VRO2022} VRO expects that software mitigations will not effectively deal with all systematic effects and the resulting spurious event triggers, especially at low brightness. Similarly, for a full buildout of Starlink and OneWeb’s proposed $\sim$48,000 satellites, every 30-second exposure on the Large Magellanic Cloud during the southern hemisphere summer months is expected to contain at least one satellite trail.~\cite{Walker2020a} 

The impact of brief satellite glints, however, is largely unknown. Such glints are ill-characterised and will almost certainly impact astronomical studies including some of the fastest-growing research areas like time-domain astronomy. One recent example is the discovery of a faint, apparently transient object thought to represent an exotic astrophysical phenomenon,~\cite{Jiang2020} later suggested to have been caused by satellite interference with the observation.~\cite{Nir2021,Michalowski2021} These instances could soon become commonplace. 

\subsection{Monetary cost of longer integration times with elevated diffuse sky brightness}

If, as models predict, the night sky becomes brighter as a result of the proliferation of space objects, then progressively longer integration times will be required to reach any given S/N threshold. For any science program with a defined S/N requirement, rising NSB inevitably imposes a loss of efficiency that can be interpreted as an increased financial burden. 

\begin{figure}[htp]
\centering
\includegraphics[width=0.9\textwidth]{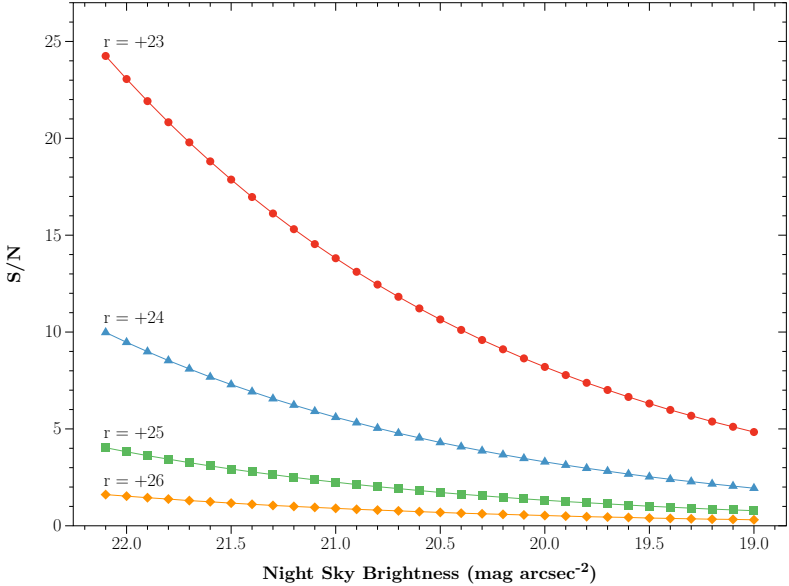}
\caption{Signal-to-noise ratio as a function of background night sky brightness for point sources in the Vera Rubin Observatory's Legacy Survey of Space and Time around the single-visit detection limit (r=+24.5 mag) and an exposure time of 20 s from predictions based on the model of Ivezić et al.~\cite{Ivezic2019} Four sources are plotted with magnitudes $r$ = +23 mag (red circles); $r$ = +24 mag (blue triangles); $r$ = +25 mag (green squares); $r$ = +26 mag (orange diamonds). Note that the numerical values on the abscissa are reversed such that the night sky becomes brighter moving left to right along the axis.}\label{Fig1}
\end{figure}

As an example, we consider the potential efficiency loss for the LSST. The survey has specified a number of quality metrics tied to S/N. Following on the LSST 5$\sigma$ detection depth quoted by Ivezić et al.~\cite{Ivezic2019}, we examined the effect of changing NSB on the LSST point-source efficiency near the single-visit detection threshold for a fixed exposure time. Figure~\ref{Fig1} shows the expected S/N as a function of the brightness of the sky adjacent to a science target. 

Given the projection by Kocifaj et al.~that zenith luminance due to space objects may approach a maximum value of 25 $\mu$cd m$^{-2}$ for solar depression angles of ~23$^{\circ}$ by 2030,~\cite{Kocifaj2021} the zenith might by then be ~12\% brighter than an assumed pristine night sky ($\sim$22.0 $m_{V}$ arcsec$^{-2}$, or $\sim2\times10^{-4}$ cd m$^{-2}$). For a sky-dominated observation near the point-source detection threshold, this 12\% increase in sky brightness directly translates to a 12\% increase in exposure time required to reach the same S/N as under pristine conditions. For brighter objects, this translation becomes less linear and the value of this factor decreases until it reaches unity (i.e., until the signal is object-dominated rather than sky-dominated).

Ivezić et al.~note that for galaxies with an i magnitude around +25, LSST photometry is expected to achieve an rms accuracy $\sigma$/(1 + $z$) of 2\% over the range 0.3 $<$ $z$ $<$ 3.0 at a S/N of 20.~\cite{Ivezic2019} To reach that S/N in a single observation would take about 12\% more exposure time for LEO space objects contributing 25 $\mu$cd m$^{-2}$ of sky brightness versus a scenario in which they contributed no sky brightness. It is important to note that this is a small effect compared to the contribution of the natural night airglow, which is typically three times brighter under quiescent conditions.~\cite{Sternberg1972} To the extent that observations of faint objects are almost always limited by the total sky brightness, any increase above the natural background set by airglow and other phenomena necessarily requires a longer exposure time to achieve a given S/N. The contribution of space objects to the diffuse sky background limiting such observations is non-negligible.

\begin{figure}[tbh]
\centering
\includegraphics[width=0.9\textwidth]{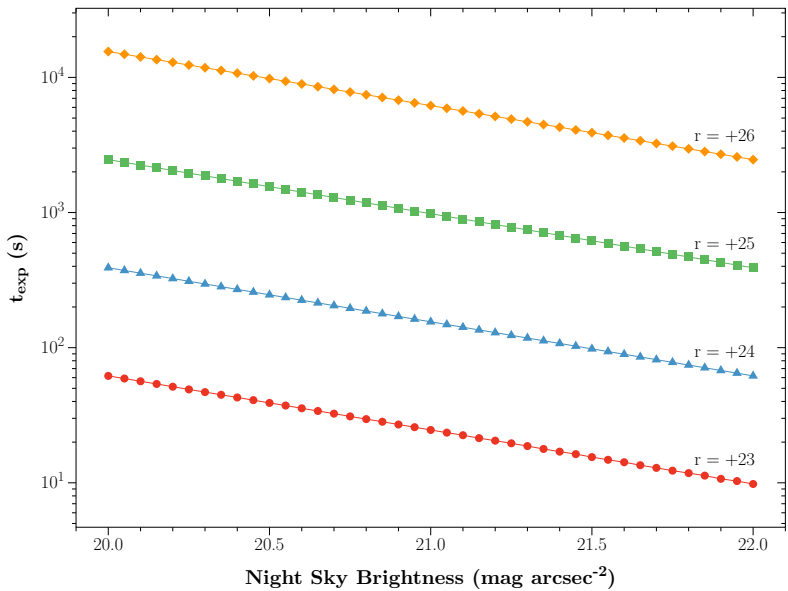}
\caption{Exposure time required to reach a S/N of 10, plotted on a log$_{10}$ scale, as a function of background NSB for point sources in the LSST. The circumstances of the plot are otherwise the same as in Figure 1.}\label{Fig2}
\end{figure}

For sky-limited photometric measurements of point sources, the exposure time ($t_{\textrm{exp}}$) required to reach a particular S/N ratio is proportional to the square of that ratio 

\begin{equation}
t_\textrm{exp} = \left(\frac{S/N}{F_{\textrm{obj}}}\right)^{2}n_{\textrm{pix}}F_{\textrm{sky}},
\end{equation}
\vspace{0.5cm}

\noindent
in which $F_{\textrm{obj}}$ is the object flux in photons s$^{-1}$, $n_{\textrm{pix}}$ is the number of detector pixels over which the point-spread function is sampled, and $F_{\textrm{sky}}$ is the sky background flux in photons s$^{-1}$ pixel$^{-1}$. For any particular combination of S/N, $F_{\textrm{obj}}$ and $n_{\textrm{pix}}$, the ratio of $t_{\textrm{exp}}$ at any two different values of $B_{\textrm{sky}}$ is constant (Figure~\ref{Fig2}): 

\begin{equation}
\frac{t_{\textrm{exp}}(B_{\textrm{2}})}{t_{\textrm{exp}}(B_{\textrm{1}})} = 10^{-0.4(B_{\textrm{1}}-B_{\textrm{2}})}
\end{equation}
\vspace{0.5cm}

\noindent
where $B_{\textrm{1}}$ and $B_{\textrm{2}}$ are arbitrary sky brightness values in units of magnitudes per square arcsecond. Furthermore, that ratio is very nearly identical to the ratio of $B_{\textrm{sky}}$ to an assumed `pristine' value at the zenith. In other words, if the NSB increases by a factor of $M$ over pristine conditions, then the exposure time required to reach an arbitrary S/N ratio at that sky brightness also increases directly by about a factor of $M$. We acknowledge that these estimates apply only to single observations and are not representative of the cumulative result of many observations. Nevertheless, given that telescope time is represented by a constant cost per unit at most facilities, we draw attention to the sobering reality that the financial cost of acquiring data under brightening night-sky conditions will scale with increasing NSB. Additionally, as annual allotments of observing time are usually a fixed quantity, if each target observing program requires more time with increasing NSB, fewer scientific programs can be completed.

Estimating the financial cost of this effect is difficult given that the relationship between survey benchmarks and operations costs is complex. The base capital cost of LSST is estimated by the U.S. National Science Foundation as \$473 million.~\cite{NSF2020} The estimated annual operation costs add up to \$290 million over the ten-year lifetime of the baseline survey. Assuming a 1:1 correspondence between survey duration and financial cost, an increase in the duration of 12\% (1.2 years) would correspond to an additional \$34.8 million in total project cost.

As a matter of fact not all objects contribute to the diffuse brightness of the sky when viewed through a telescope. Bright objects appear in images as discernible streaks. In order to appear as a streak, the irradiance of an object image on the detector plane must be sufficient to provide a detectable signal above the background and noise levels. Recall that the image on the detector is the two-dimensional convolution of the geometrical image of the object (whose typical linear dimension is proportional to the object size and inversely proportional to the object distance) and the overall point spread function of the optical system (determined by the optical response of the telescope and the relevant atmospheric effects averaged within the effective exposure time). The signal of the object on a pixel is proportional to the irradiance of its image, the pixel area and the effective exposure time (the time during which the moving image of the object illuminates the pixel’s active surface). The radiance contribution of the objects detectable as individual streaks shall be subtracted from the total diffuse radiance mentioned above. For facilities such as the Rubin Observatory, whose telescope and camera should detect streaks from objects in orbit potentially as small as $\sim$5 cm, this would remove about three parts in eight of reflected sunlight from the maximum 25 $\mu$cd m$^{-2}$ that would otherwise elevate the natural diffuse sky background. In practical terms this means that in these conditions the diffuse background will be (5/8) × 25 $\mu$cd m$^{-2}$ $\simeq$ 15 $\mu$cd m$^{-2}$; that is, a 7.5\% increase over the assumed natural reference level. Considering this effect alone, the LSST project would therefore require 7.5\% more time, and at equal yearly cost this would be an overcharge of 0.075 $\times$ \$290 million = \$21.8 million.

\subsection{Shortcomings of the current modelling regime}

We do not have a way to easily predict or calculate the time evolution of the number of space objects as a function of their sizes and orbit altitude distribution. Instead, presently we have to live with the aftermath of unplanned events or debris cascades. As Kocifaj et al.~point out, the most complete picture of the cumulative number of objects in LEO larger than a given size dates to the mid-1990s.~\cite{Kocifaj2021} The international scientific community is limited to publicly available, civilian data typically derived from limited radar measurements~\cite{Muntoni2021} or space environment exposure experiments;~\cite{Mandeville1995} we speculate that militaries and private companies may have access to more detailed information. Given the high relative velocities involved, the still largely unknown population of mm-sized and smaller objects poses the greatest threat to space hardware. The hazard is expected to be significant for spacecraft in large satellite constellations.~\cite{LeMay2018}

Furthermore, there is great variety, and therefore uncertainty, in the number and type of satellites proposed for near-term orbital deployment. Recently launched objects are observed to vary significantly in terms of their optical properties, both between individual objects and for a single object over time.~\cite{Mallama2021a,TregloanReed2021,Mallama2022} Individual debris fragments are even less well-understood, although their properties in aggregate can be estimated reasonably well. But the single greatest uncertain factor in the modelling realm, and most challenging to predict, is what exactly will dominate the changing NSB. We do not yet know whether it will be the over 400,000 satellites already planned for launch in the coming decade;~\cite{McDowell2023} a cascade of fragments from accidental collisions; debris from reckless acts undertaken in LEO by state actors; or some combination of these.

\section{Envisioning the broader landscape of consequences}\label{sec4}

\subsection{Loss of dark night skies}

Brightening night skies impact not only astronomy but also the human experience of viewing the night sky. We attempt to quantify this through comparing viewing experiences under various sky brightness conditions. There are a number of ways of characterising the brightness of the night sky using both qualitative and quantitative metrics.~\cite{Barentine2022a} In Table~\ref{Table1} we compare several key visual night sky quality indicators for a series of levels of sky brightness above the ‘pristine’ brightness assumed above, assuming a dark-adapted observer viewing the night sky under typical clear-sky conditions at sea level. It also makes the Milky Way significantly harder to see from anywhere on Earth, and diminishes views of night sky phenomena such as night airglow, weak aurorae, and faint meteors. Doubling the zenith brightness relative to an unpolluted night sky reduces the number of visible stars by roughly 30\% and reduces the number of visually observable meteors by a factor of up to one half.

\begin{table}[thp]
\begin{center}
\caption{Visual night sky quality indicators as a function of rising zenith brightness over the range of values from pristine conditions to those at which the zenith is twice as bright in comparison. The table notes are as follows: (1) from Equation 53 in~\cite{Crumey2014} for a typical observer (field factor F = 2); and (2) a qualitative scale ranking night skies from 1 (pristine) to 9 (most light-polluted) based on a series of visual criteria.~\cite{Bortle2001}}\label{Table1}%
\begin{tabular}{@{}ccccc@{}}
\toprule
\bf{Zenith} 					& \bf{Per cent} 			& \bf{Limiting}	 	& \bf{} & \bf{Number of}\\
\bf{Brightness}					& \bf{increase over} 		& \bf{Visual}	 	& \bf{Bortle}	& \bf{visible stars}\\
\bf{($m_{V}$ arcsec$^{-2}$)} 		& \bf{pristine night sky}	& \bf{Magnitude\footnotemark[1]} 	& \bf{Scale\footnotemark[2]} & \bf{per night}\\
\midrule
22.00 & 0	& +6.3 & 1 & 3500 \\
21.99 & 1 & +6.3 & 1 & 3500 \\
21.97 & 3 & +6.2 & 1 & 3150 \\
21.95 & 5 & +6.2 & 1 & 3150 \\
21.90 & 10 & +6.2 & 1 & 3150 \\
21.76 & 25 & +6.2 & 2 & 3150 \\
21.56 & 50 & +6.1 & 3 & 2800 \\
21.25 & 100 & +6.0 & 4 & 2540 \\
\botrule
\end{tabular}
\end{center}
\end{table}

\subsection{Impacts on human heritage and culture}

The night sky transcends science or utility; it is equally a source of inspiration, connection to nature and recreation. For some cultures, sky traditions are a prominent aspect of their social customs, cultural traditions and religious beliefs. As such, it represents a form of intangible human heritage that deserves intentional preservation and safeguarding for future generations.~\cite{Ruggles2010,Venkatesan2020}

Our calculations indicate that the brighter stars and constellations often utilised in navigational aspects of cultural sky traditions, including wayfinding, will remain visible even for the more extreme scenarios we consider here. However, the anticipated rise in NSB adds to the contribution of terrestrial skyglow and will wash out fainter stars and the Milky Way. This tends to diminish visibility of the dark clouds seen in silhouette against the Milky Way that play an important role in many sky cultural traditions in the Southern Hemisphere.~\cite{Gullberg2020} Fainter objects, such as nebulae, star clusters and dimmer groups of stars are also often key elements of teachings in various Indigenous communities, as are observations of the heliacal rising of various celestial objects.~\cite{Lee2020} These are all likely to be impacted by rising NSB. In addition, satellites visible as moving points of light alter the appearance of the night sky; for some communities, this is seen to ``interrupt'' their ``relationship with the stars and ceremonial ways of connecting with them''.~\cite{Venkatesan2021} While various mitigation techniques may help address some satellite constellation impacts to professional astronomy, real-time observations, and thus living sky traditions, will be adversely impacted by visible satellites and rising NSB.

\subsection{Implications for equity, inclusion and accessibility in astronomy}

The roiling waves of the COVID-19 pandemic, climate change, and global economic turbulence in recent years have resonantly combined in unpredictable ways to jeopardise the lives and livelihoods of the world's most vulnerable populations. Significant adverse health effects also arise from disproportionate light and noise pollution in these communities.~\cite{Nadybal2020} Rising NSB has a documented impact on the health of humans and the wellbeing of broader ecological systems;~\cite{Venkatesan2021} this is especially concerning given its inescapable, planet-wide nature. Reduced visibility of, e.g., the Milky Way impacts migratory patterns of many creatures;~\cite{Venkatesan2021,Foster2018,Stone2015,Pakhomov2017} Lawrence et al.~point out that most circadian rhythms are apparently controlled by diffuse ambient light and not by moving point sources.~\cite{Lawrence2022}

For professional astronomy, in this time of shrinking budgets and fewer grant dollars in a zero-sum game, the competition for observing time on ground-based telescopes and facilities will become even more highly competitive than it is at present, especially if longer exposures are needed for sources in a sky with greater radiance from space objects. Observing time, grants and awards, like all privilege, tend to accumulate in select academic lineages and institutional classes; longer integration times are likely to even further concentrate this privilege within a shrinking circle of institutions. In such a professional environment, recruiting, retention and promotion of underrepresented and marginalised groups in astronomy faces increasing challenges, at a time when our field's future workforce is already confronting a radically altered landscape of professional and research opportunities.~\cite{Venkatesan2020}

The pandemic has revealed many pre-existing systemic conditions that have widened socioeconomic and learning gaps, including access to affordable global broadband.~\cite{Patrick2021} Very few would argue against the dire need for expanded access to broadband; our capacity to conduct teaching and research in astronomy, as well as a competitive future workforce in astronomy and other fields, depends on this. While the commercial space industry has argued for the benefits of providing broadband Internet from space, we note that, to date, no company has ever demonstrated a satellite broadband business model that is both profitable and sustainable. The latter is often mentioned prominently, but the driving factor is the former: profit. Rawls et al. questioned the motive often stated by commercial space companies that have proposed launching megaconstellations, which is to provide broadband to underserved populations globally; instead, they found that ``the Internet service offered by these satellites will primarily target populations where it is unaffordable, not needed, or both.''~\cite{Rawls2020}

Furthermore, the effective consolidation of control of LEO space for communications by a handful of private companies, or by a small number of privileged industrialised states, risks diminishing the equity, inclusion and accessibility to broadband communications. We note that technological alternatives to broadband from orbit exist, such as fibre-optic-transmission and the latest generation of terrestrial wireless data networks, that could achieve the same result without the scientific and business risks attendant to the launch and operation of satellite megaconstellations. In addition, distributed broadband ground communication networks are somewhat more difficult to oligopolise.

\section{Mitigations: potential gains and risks}\label{sec5}

Mitigating the threats to astronomy described above largely falls into two categories: (1) modifying satellite and satellite constellation designs, and (2) back-correction or restoration of astronomical data impacted by satellites and space debris.~\cite{Massey2020,Walker2020a,Lalbakhsh2022} The former approach was taken by SpaceX in 2020 after astronomers first raised the prospect of harm to astronomy research resulting from its Starlink satellites. Its `DarkSat' and `VisorSat' designs were intended to reduce the brightness of Starlink objects. Some observers measured decreases in on-station brightnesses among objects of these designs by up to 1.5 magnitudes compared to the original Starlink design,~\cite{Cole2020a,TregloanReed2020,TregloanReed2021,Mallama2021b} while others saw essentially no change~\cite{Cole2020b} or even brightness increases.~\cite{Horiuchi2020} However, both mitigations were abandoned for engineering reasons, and neither achieved the goal of reducing the brightness of first-generation Starlink objects at station to below the threshold of unaided-eye observability. Although this problem remains unsolved, SpaceX and its competitors are planning to launch new satellites of unknown brightness.

Design solutions may reduce streak brightnesses but are not expected to have any meaningful effect on rising diffuse sky brightness, since most of the increase in NSB is due to small debris. Design solutions will also have little or no effect on the potential for debris-generating collisions. In addition, mitigating solutions at one bandpass may be a problem for other bandpasses; for example, optically darker objects often radiate more brightly in the infrared and submillimetre, which creates interference with ground-based observations at those wavelengths. It also remains a problem that many operators are not forthcoming with details of their satellite designs, including materials, albedos, bidirectional reflectance distribution functions (BRDFs) and other parameters, because they protect them as intellectual property and governments generally do not require their public disclosure. That often leaves astronomers to invest their taxpayer-funded time and resources into inferring these properties through observations and modelling. We note that some operators have been more collaborative with astronomers and with data sharing, e.g., SpaceX’s recent report on their attempts to decrease satellite reflectivity.~\cite{SpaceX2022} Significant concerns remain about the brightness of next-generation Starlink satellites, as well as the extraordinarily bright AST SpaceMobile BlueWalker 3 prototype satellite proved quite recently.~\cite{Mallama2022}

Data treatments can improve the quality of images impacted by trails and glints and recover some science pixels. But saturated pixels are lost, and they can induce crosstalk between detector amplifiers. In their simulations of LSST data, Tyson et al.~found that they could effectively remove the influence of nonlinear image artefacts induced by satellite trails at a maintained brightness of magnitude +7 or fainter, although they expected that “systematic errors that may impact data analysis and limit some science'' would remain after correction.~\cite{Tyson2020}

Perhaps the mitigation least palatable to the commercial space industry, but one that governments may certainly impose, is to simply launch fewer satellites into near-Earth space. It is the only solution that simultaneously tackles all the problems we describe here. With any such plan for satellite launch reduction should come the intent to responsibly de-orbit those already launched at end-of-mission (EOM) in order to minimise ongoing collision risks. Satellite operators should also be held to broader standards of responsible reduction, and ideally elimination, of debris associated with all stages of satellite operation including launch and disposal at EOM.~\cite{Byers2022} While U.S. policy was recently amended to reduce the disposal time from within 25 years to within five years of EOM as a condition of licensing spacecraft operations,~\cite{FCC2022} not all jurisdictions make the same demand for objects launched from their territories. On-orbit debris removal has been proposed,~\cite{Shan2016,Mark2019} but to date no one has demonstrated this successfully in practice. The same is true for perfect collision avoidance through various space situational awareness schemes.~\cite{Rybus2018,Oltrogge2019} It is a far more reliable course of action to avoid debris cascades in the first place, which necessarily entails launching fewer objects and reducing the number of objects already in orbit.~\cite{Lewis2020}

This brewing crisis in LEO has powerful lessons for our shared future in near-Earth space. There are still opportunities to get ahead of the problem elsewhere. For example, now is the time to consider a future in which astronomical observations performed from the lunar surface may be similarly affected by a growing swarm of space objects orbiting the Moon.~\cite{Silk2020} There are fewer legal restrictions on the use of cislunar space, and the race to occupy that space is already on.~\cite{Spudis2011}

\section{Looking ahead}\label{sec6}

The lack of coordinated and effective global policy, regulation, and oversight among spacefaring nations and space actors has led to the prospect of hundreds of thousands of satellites planned for launch in the coming decade, without attendant conditions on effective mitigation strategies or environmental self-assessment as a condition of licensing. Despite a narrative of democratising space and delivering affordable global broadband, it is a model that prioritises urgency, privatised benefits and short-term goals over real sustainability and the public interest. This also ignores our shared ancestry and heritage in space. 

In this paper, we have emphasised the potential consequences resulting from a global increase in sky brightness given the rising numbers of space objects in LEO. Unlike satellite streaks and glints that hold some options for back-corrections, this is an inescapable, planet-wide phenomenon that will affect professional and casual observations of the skies, as well as myriad biological systems. We have attempted some estimates of the loss of professional observing time and unaided-eye observations of astronomical phenomena and pointed to the work of other teams on space environmentalism and the risks from space debris.

We also honour that the broader loss of dark skies for humanity is essentially incalculable, given the many ways we have connected to the night sky for millennia. The night sky needs preserving and defending for future generations who may not know what we have today, akin to the analogous situation of disappearing rainforests and glaciers. But unlike these examples of the toll of climate change and human activity, we are still in the early stages of changing the night sky and the environment of space for future generations. Many of the consequences of this remain unknown. 

To paraphrase the Senegalese conservationist Baba Dioum, we will conserve only what we value, value only what we know, and know only what we are taught. We conclude by offering the hope that we can still deliver on the promise of human activity and development in and from space by proceeding sustainably for all stakeholders and preserving space as a resource for future generations. We must become proactive now or risk irreparable harm to astronomy resulting from the current rush to capitalise on the exploitation of LEO and other space resources. 

\backmatter

\bmhead{Acknowledgments}

M.K. acknowledges support from the Slovak Research and Development Agency under contract number APVV-18-0014. A.V. gratefully acknowledges support from the University of San Francisco Faculty Development Fund.

\bmhead{Competing Interests}

Four authors (J.B., A.V., J.H. and J.L.) are unpaid members of committees of the American Astronomical Society and International Astronomical Union whose scope of concern includes the topics covered by this article. The views expressed in this article do not necessarily represent the positions of either organisation. 

\bibliography{barentine-et-al-2023}


\end{document}